\documentclass{IEEEtran}  

\IEEEoverridecommandlockouts                              

\makeatletter
\renewcommand{\@IEEEsectpunct}{.\ \,}
\makeatother

\usepackage{graphics} 
\usepackage{epsfig} 
\usepackage{mathptmx} 
\usepackage{times} 
\usepackage{amsmath} 
\usepackage{amssymb}  
\usepackage[]{units}
\usepackage{gensymb}
\usepackage{placeins}
\usepackage{dblfloatfix}  
\usepackage{arydshln}
\usepackage{xcolor}

\title{\LARGE \bf
A Deep Matched Filter For R-Peak Detection in Ear-ECG}

\author{Harry J. Davies, Ghena Hammour, Marek Zylinski, Amir Nassibi and Danilo P. Mandic\\
(harry.davies14, d.mandic)@imperial.ac.uk
}

\begin{document}

\maketitle
\thispagestyle{empty}
\pagestyle{empty}

\begin{abstract}

The Ear-ECG provides a continuous Lead I electrocardiogram (ECG) by measuring the potential difference related to heart activity through the use of electrodes that can be embedded within earphones. The significant increase in wearability and comfort afforded by Ear-ECG is often accompanied by a corresponding degradation in signal quality - a common obstacle that is shared by the majority of wearable technologies. We aim to resolve this issue by introducing a Deep Matched Filter (Deep-MF) for the highly accurate detection of R-peaks in wearable ECG, thus enhancing the utility of Ear-ECG in real-world scenarios. The Deep-MF consists of an encoder stage (trained as part of an encoder-decoder module to reproduce ground truth ECG), and an R-peak classifier stage. Through its operation as a Matched Filter, the encoder section searches for matches with an ECG template pattern in the input signal, prior to filtering the matches with the subsequent convolutional layers and selecting peaks corresponding to true ECG matches. The so condensed latent representation of R-peak information is then fed into a simple R-peak classifier, of which the output provides precise R-peak locations. The proposed Deep Matched Filter is evaluated using leave-one-subject-out cross validation over 36 subjects with an age range of 18-75, with the Deep-MF outperforming existing algorithms for R-peak detection in noisy ECG. The proposed Deep-MF is bench marked against a ground truth ECG in the form of either chest-ECG or arm-ECG, and both R-peak recall and R-peak precision is calculated. The Deep-MF achieves a median R-peak recall of 94.9\% and a median precision of 91.2\% across subjects when evaluated with leave-one-subject-out cross validation. Moreover, when evaluated across a range of thresholds, the Deep-MF achieves an area under the curve (AUC) value of 0.97. The interpretability of Deep-MF as a Matched Filter is further strengthened by the analysis of its response to partial initialisation with an ECG template. We demonstrate that the Deep Matched Filter algorithm not only retains the initialised ECG kernel structure during the training process, but also amplifies portions of the ECG which it deems most valuable - namely the P wave, and each aspect of the QRS complex. Overall, the Deep Matched Filter serves as a valuable step forward for the real-world functionality of Ear-ECG and, through its explainable operation, the acceptance of deep learning models in e-health. 

\end{abstract}


\section{Introduction}

\IEEEPARstart{R}{ecent} advancements in Hearables serve to disrupt the e-health market through the provision of continuous monitoring of mental state and vital signs from the ear \cite{Goverdovsky2017}. Of the different Hearables' sensing modalities, one of the most notable is the Ear-ECG, which provides continuous lead I electrocardiogram (ECG), the measurement of the electrical activity of the heart, through the potential difference between two in-ear electrodes on separate sides of the head \cite{Von_Rosenberg2017}. The precise heart rate information from ECG can be used to monitor stress through heart rate variability metrics \cite{Adjei2019}\cite{Hye_Geum_2018}\cite{Pecchia2011} and the detection of irregular heart rhythms (arrhythmia) \cite{Hammour2019}. However, with the immense gain in comfort and wearability afforded by an in-ear sensor compared to electrodes on the chest, comes a drop in signal to noise ratio. The potential difference across the heart is often as much as 2 orders of magnitude lower from the ear than it is at the chest \cite{Yarici2022}. Moreover, the Ear-ECG commonly contains other signals comparable in amplitude, such as electrical activity generated by eye movements, known as electrooculography (EOG) \cite{Skoglund2022}, and electrical signals generated by neuronal activity in the brain, known as electroencephalography (EEG) \cite{Goverdovsky2016}\cite{nakamura2020}\cite{Kidmose2013}. In order to best exploit the benefits of Ear-ECG, algorithms need to be able to detect the presence of ECG waveform across challenging range of signal qualities, and correctly distinguish the peaks in ECG (R-peaks) from peaks that may occur due to artefacts or other electrical activity.

\begin{figure}[b]
\centerline{\includegraphics[width=0.5\textwidth]{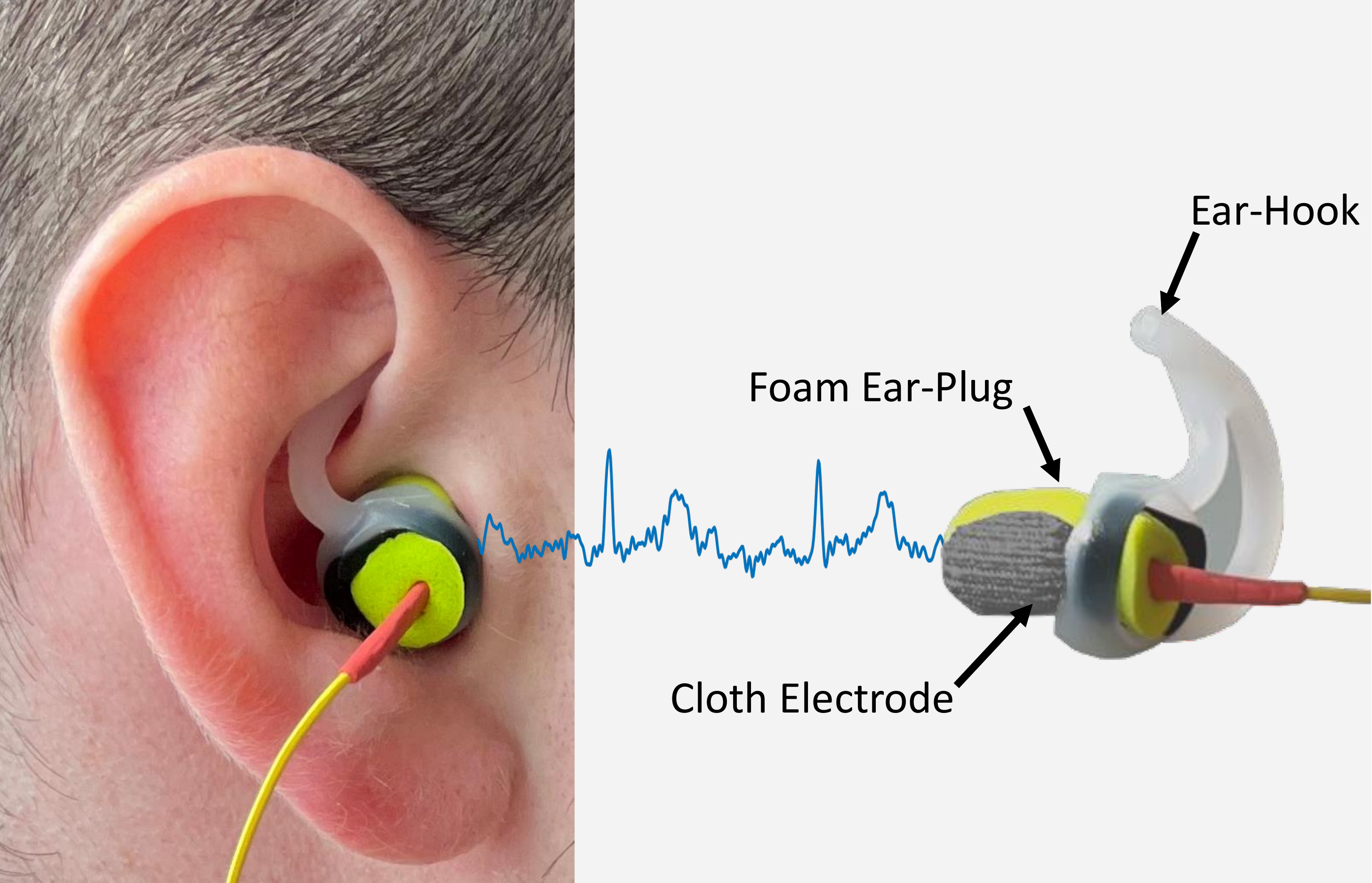}}
\caption{The Ear-ECG earpiece. Left: The placement of one of the ear electrodes within the ear canal. Right: A labelled prototype Ear-ECG device, consisting of a foam ear-plug, a cloth electrode and an ear-hook to stabilise the ear-piece within the ear canal.}
\label{hardware}
\end{figure}

\begin{figure*}[h!]
\centerline{\includegraphics[width=\textwidth]{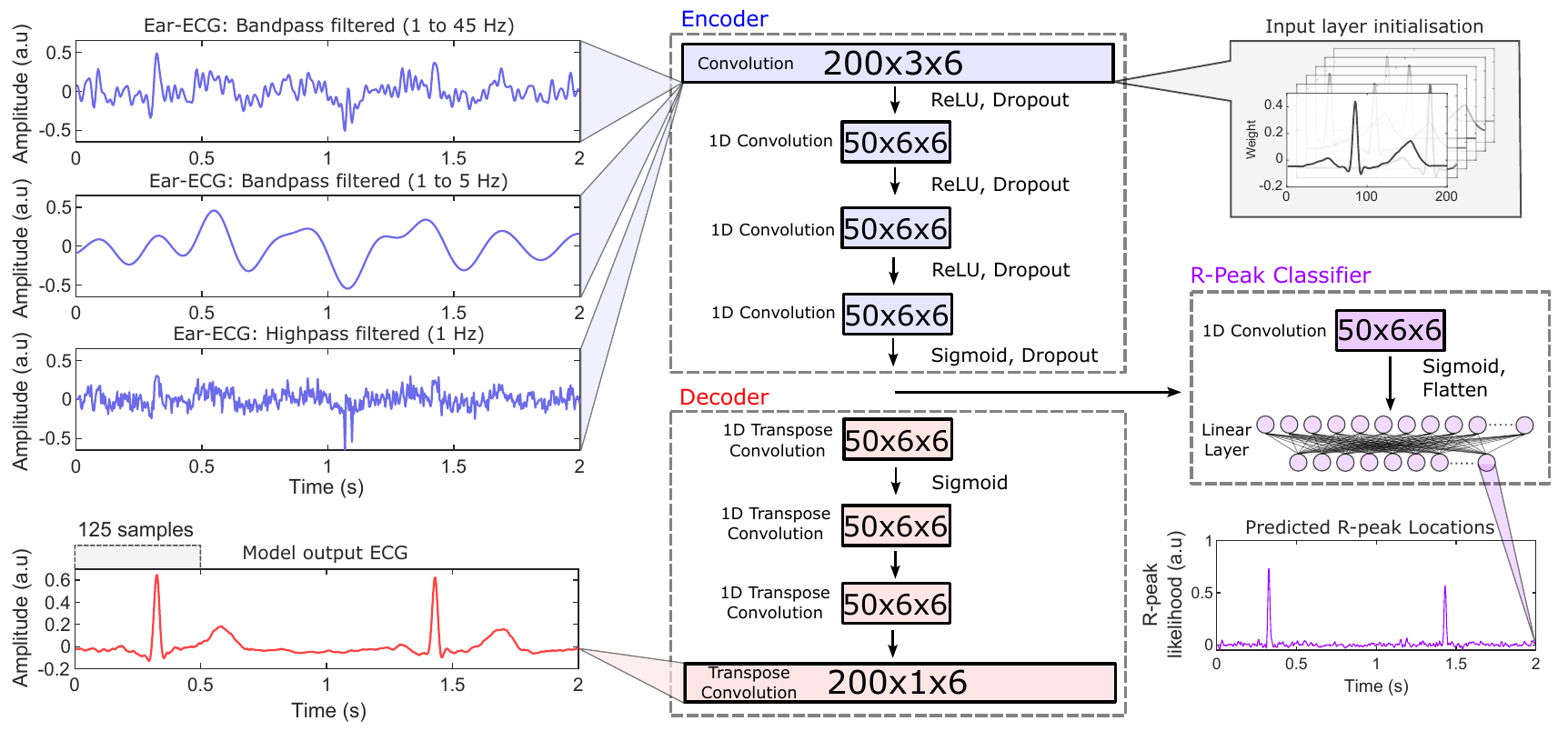}}
\caption{An overview of the proposed Deep Matched Filter (deep-MF) architecture. The three input channels to the model (top left, blue), including Ear-ECG band-pass filtered between 1 and 45Hz, Ear-ECG band-pass filtered between 1 and 5Hz, and Ear-ECG high-pass filtered with a cut-off frequency of 1Hz. These channels are inputs to an encoder module which serves as a matched filter (middle, blue). The encoder, constructed of 1D convolutional layers, consists of a matched filter layer with kernels of length 200 (0.8 seconds) which serve to detect ECG patterns in the input, and three subsequent ``refinement" layers with kernels of length 50, to determine which matches are true. A subsection of the weights of the matched filter layer are initialised with a shifted ECG template (top right, grey). The encoder is accompanied by a decoder (middle, red), consisting of 1D transpose convolutional layers, which upsample the output of the encoder into an output which resembles an ECG waveform (bottom left, red). The decoder is essential for the training of the encoder. The final module is the R-peak classifier (right, purple), which takes the output of the encoder and uses it to predict the position of the R-peak. The R-peak classifier consists of a single 1D convolution layer, and a linear layer.}
\label{deep_learning_architecture}
\end{figure*}

With this in mind, it is straightforward to assume that a matched filter \cite{Turin1960}, the process of shifting a template across a signal to enhance the detection of the pattern contained within the template, would perform well in the scenario of detecting R-peaks in wearable ECG. This has been demonstrated previously through the combination of matched filter and Hilbert transform \cite{Chanwimalueang2015}, which was shown to outperform the commonly used Pan-Tompkins algorithm \cite{Tompkins1985} for r-peak detection. Recent work on the interpretability of convolutional neural networks (CNNs) has demonstrated that at a fundamental pattern recognition level, a CNN performs in the same way as a matched filter, by performing convolution between a learned template kernel and an input signal or image and exploiting the correlation between the two \cite{Stankovic2023}. This was further verified through the MNIST handwriting data set, in which trained kernels converged to resemble different numbers \cite{li2022}. Given the clear benefits of using matched filtering to detect R-peaks in noisy ECG, and the theoretical link between CNNs and matched filtering, it is hypothesised that a learned convolutional matched filter could be leveraged to provide superior results for R-peak detection, whilst remaining fully interpretable in its operation.  

To this end, we implement a deep convolutional neural network based matched filter for the efficient and accurate detection of R-peaks in Ear-ECG with poor signal to noise ratio. The trained model, whilst demonstrating exceptional performance over existing methods, has the benefit of full interpretability through the lens of matched filters, with kernel weights that exploit and amplify aspects of the ECG pattern.

\section{Methods}

\subsection{Hardware and Data}

Simultaneous Ear-ECG and either arm-ECG or chest-ECG (resembling lead I) was measured from 36 subjects, with an age range of 18-75. There was a minimum of 2 minutes of data recorded from each subject, with the majority of subjects having 5 minutes of ECG data. Recordings took place when subjects were still or during sleep to minimise the impact of motion artefacts, but it should be noted that motion artefacts were still present in the data, albeit rare, and not excluded from our analysis. In 34 of the subjects, the Ear-ECG was recorded with two earpieces across the head with a ground electrode placed on the forehead. In 2 of the subjects the Ear-ECG signal was from a single ear electrode which was referenced to the contra-lateral mastoid. The Ear-ECG earpiece, shown in Fig.~\ref{hardware}, consisted of a foam earpiece with a cloth electrode, and electrode gel was used to reduce the impedance between the electrodes and the skin of the ear canal. The  recordings  were  performed  under  the  IC  ethics  committee  approval  JRCO  20IC6414. All  subjects  gave  full informed consent.

The Ear-ECG was down-sampled from 500Hz to 250Hz, and pre-filtered with three separate configurations to provide 3 input channels to the model. The first channel was a band-pass filter between 1Hz and 4Hz which aimed to reduce higher frequency noise whilst preserving the crucial information in the ECG. The second channel was a band-pass filter between 1 and 5Hz, which removed high frequency noise and the QRS complex from the ECG, but retained information on the P and T waves. The third channel was high-pass filtered with a cut-off frequency of 1Hz. This preserved all of the higher frequency detail present in the ECG, but also retained high frequency noise such as electrical interference at 50Hz. To segment the data, a sliding window with a length of two seconds (500 samples) was implemented with a shift of 0.4 seconds (100 samples). This resulted in a total of 26564 segments across all subjects. Two seconds was chosen as the segment length so that inputs would always have an ECG waveform, and usually have upwards of two ECG waveforms.

\begin{figure}[h!]
\centerline{\includegraphics[width=0.5\textwidth]{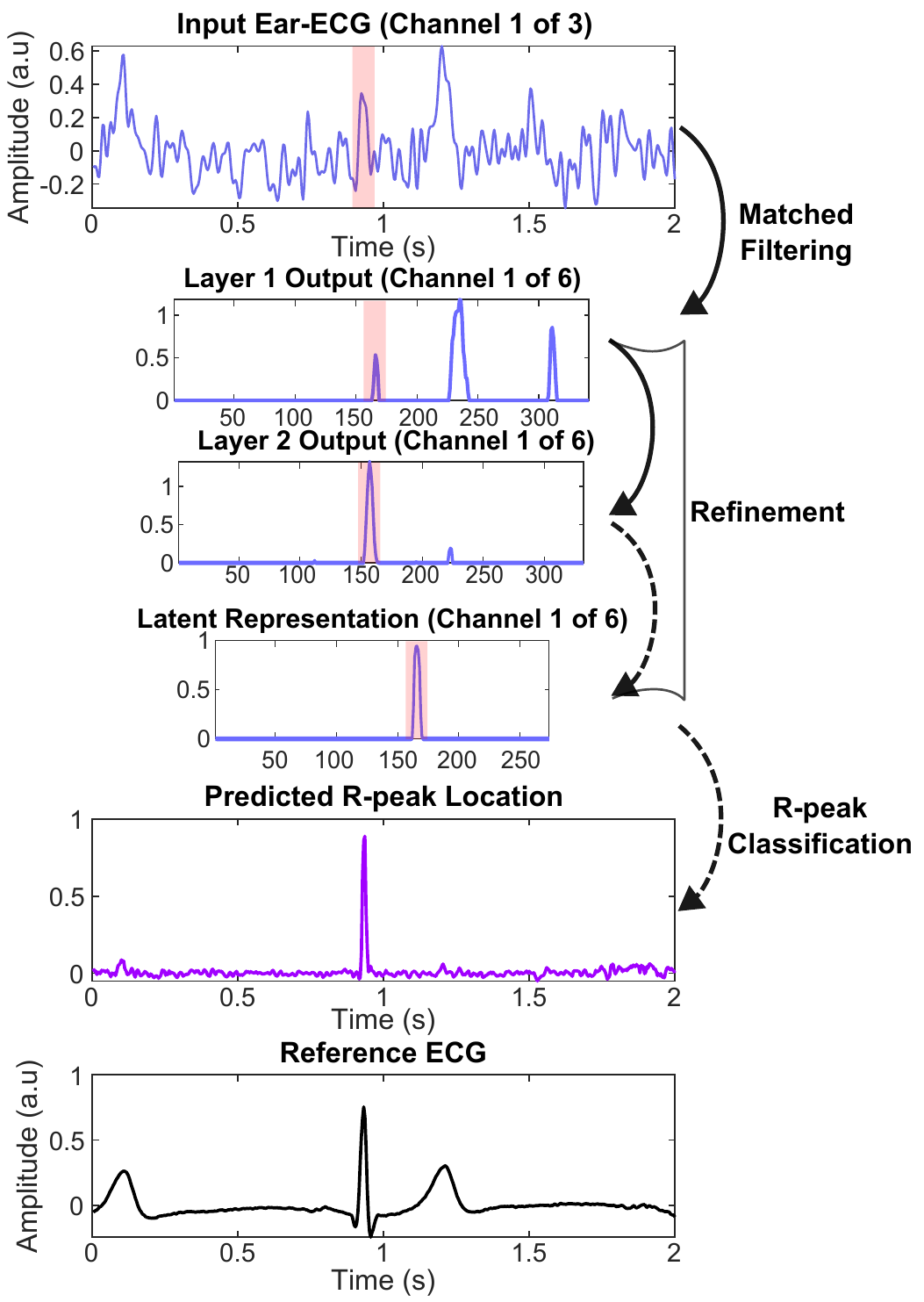}}
\caption{The signal pathway through the proposed Deep Matched Filter (Deep-MF) of an example test input Ear-ECG trace (blue). The input firstly passes through the matched filter layer, resulting in the Layer 1 Output, with 3 potential matches. This initial output is then passed through the subsequent ``refinement" layers, until a singular peak is present in the latent representation. In the matched filter and refinement stages, the true peak is highlighted with a shaded red box. This latent space is then passed through the R-peak classification phase, resulting in predicted R-peak location (purple). For the purposes of comparison, the ground truth ECG is displayed below in black.}
\label{model_walkthrough}
\end{figure}

\subsection{Deep Matched Filter Model}

The deep matched filter architecture, developed in PyTorch \cite{pytorch_cite} and shown in Fig.~\ref{deep_learning_architecture}, consists of two main parts. Firstly, an encoder-decoder module, which aims to extract shared information between the input and a training reference by condensing the information from the input that is most predictive of the output into a latent representation \cite{DAVIES2023}. In this case, the encoder-decoder was trained with arm-ECG as a reference and aims to encode the shared information between the Ear-ECG and the arm-ECG, before decoding this information into a waveform resembling that of the arm-ECG. The encoder, whilst similar in organisation to that of a denoiser, behaved as a matched filter by simply encoding the r-peak location from the original Ear-ECG and no corresponding morphological information. It then used this encoded R-peak location and pasted a learned ECG pattern in the same position. The decoder was thus bypassed, with a simple CNN based classifier which used the latent representation to predict the R-peak location. 

The encoder, highlighted in blue in Fig.~\ref{deep_learning_architecture}, consisted of 4 one-dimensional convolutional layers. In the first layer, there were 6 kernels associated with each input channel, to form 6 output channels. In all subsequent layers there were 6 kernels associated with each of the 6 new inputs. In the first layer, a kernel size of 200 was chosen, corresponding to 0.8 seconds and representing a duration slightly longer than that of a full ECG segment for its use as a matched filter template. Moreover, the 6 kernels corresponding to the first band-pass filtered input channel (1 to 45Hz) were initialised with a shifted ECG template which is highlighted in grey in Fig.~\ref{deep_learning_architecture}. The subsequent layers in the encoder had a kernel size of 50, chosen to encompass the width of the resulting "match" peak from convolution between the input and the input layer. These layers served as refinement layers for the output of the first layer, in essence helping the model to increase the precision of the matched filter by deciding which matches were valid and which matches were not. The first 3 layers had a ReLU activation function and a dropout of 50\%, and the fourth layer had a Sigmoid activation function and a dropout of 50\%. The Sigmoid activation function was important for ensuring stability of the model during training, due to the bounded output property. 

The decoder, highlighted in red in Fig~.\ref{deep_learning_architecture}, contained 4 transpose convolutional layers which mirrored the one-dimensional convolutional layers of the encoder. In contrast to the encoder layers, there was no dropout applied and only a single Sigmoid activation function was applied between the first and second decoder layers. Moreover, there was a single output corresponding to a 2 second ECG trace. The encoder-decoder structure was trained to minimise mean squared error between the output and the reference ECG waveform. Importantly however, due to the encoder operating as a matched filter and the fact that there were only slight differences in the morphology of ECG waveform across subjects, the model minimised error by detecting only the location of the ECG in the input, and upsampling this into a generic ECG waveform. Despite this, the training paradigm of using a decoder to replicate a full ECG waveform was necessary, as when the same structure was trained to replicate just the R-peaks it often failed to converge.

\begin{figure*}[h!]
\centerline{\includegraphics[width=\textwidth]{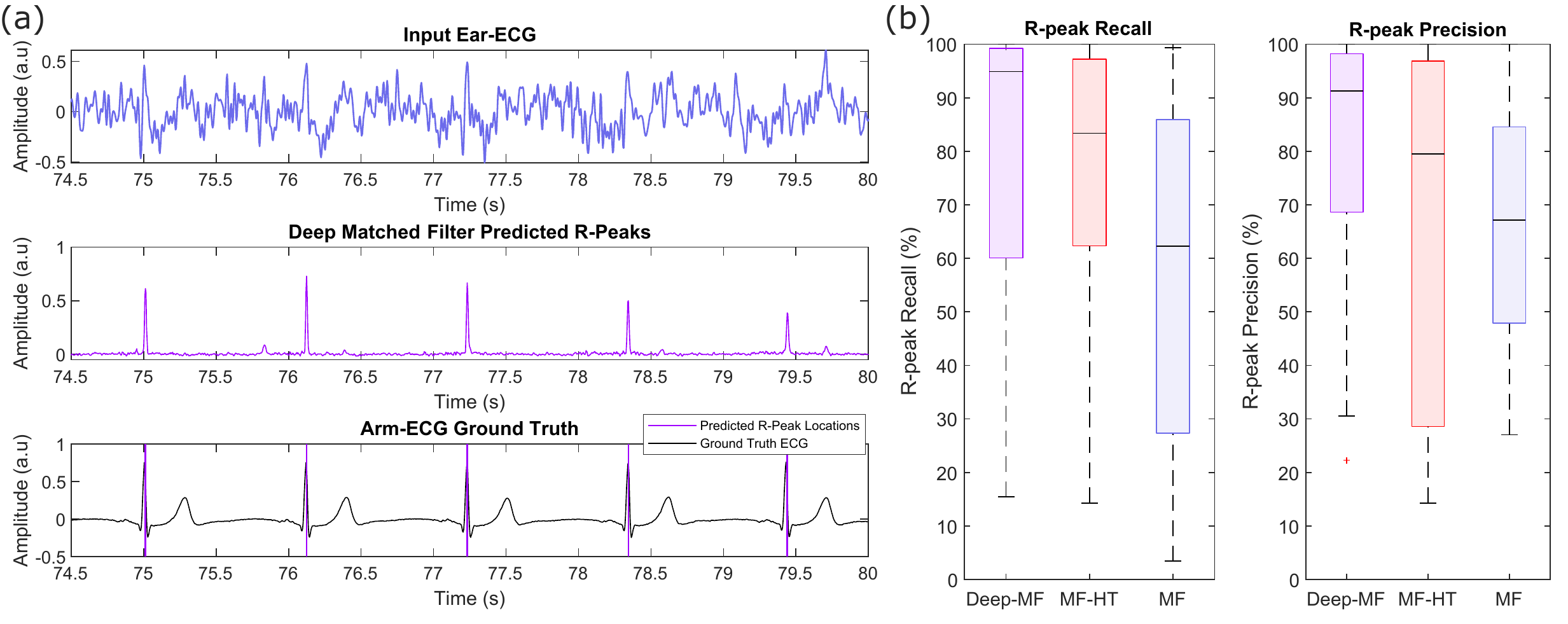}}
\caption{Test results for the proposed Deep Matched Filter for R-peak detection in noisy Ear-ECG. (a) An example of Ear-ECG with a poor signal to noise ratio (blue) and the corresponding output R-peak locations of the Deep-MF (purple). Below is the ground truth Arm-ECG (black) with predicted R-peak locations overlaid in purple. Note that in the input there are several peaks which a stronger than the true R-peaks, particularly between 79 and 80 seconds. The Deep-MF correctly rejects these peaks and predicts the true R-peaks in the output. (b) Boxplots of R-peak Recall and Precision, across all subjects as a result of leave one subject out cross validation. The results of the proposed Deep-MF filter (purple) are compared to the Matched Filter Hilbert Transform (MF-HT) (red) and the standard Matched-Filter (blue). In terms of R-peak recall, the percentage of peaks in the ground truth correctly identified by the model, the proposed Deep-MF achieves a median of 94.9\%, compared with the MF-HT and MF where the respective median recalls are 83.4\% and 62.3\%. In terms of precision, the percentage of the peaks predicted by the model that are correct, the proposed Deep-MF achieves a median of 91.2\%, compared with the MF-HT and MF which achieved the respective median precisions of 79.5\% and 67\%.}
\label{results_mfht}
\end{figure*}

Given that the latent representation contained information on location of the ECG in the input, a simple classifier, highlighted in purple in Fig.~\ref{deep_learning_architecture}, was then trained to take in the latent variables and output R-peak locations. This classifier consisted of a single layer 1D convolution, followed by a Sigmoid activation function and flattening, before finally being passed to a linear layer. For the training of this second model, the latent variables were extracted from the training inputs being passed through the trained encoder. These latent variables were then used as inputs to the R-peak classifier which was trained against the corresponding R-peak locations from the reference ECG. The R-peak locations were calculated from the ECG reference using the MATLAB (ver. 2022b) function findpeaks, and both the location of the R-peak and the two neighbouring values were assigned a value of 1. Extending the window of the R-peak location from 1 to 3 in the training reference gave the model slightly more lenience in the shift of a peak, and without this the model had a tendency to suppress peaks. The output of the classifier was trained to minimise mean squared error against the corresponding array of 1s and 0s. Finally, averaging was performed on the 2 second output of the deep matched-filter, with a shift of 0.4 seconds. In a real-world setting it would be practical to implement the model with a rolling output, rather than waiting for each new 2 second window to pass. Moreover, if the ECG in the input was at the boundaries of the model and not a full waveform it would be cropped with respect to the matched filter (an issue that padding would not solve) and thus it would be more difficult to detect. This issue is circumvented by using a rolling window by ensuring that every ECG waveform in the input is at some point close to the center of the input.  

The encoder-decoder model was trained for 10 epochs and the R-peak classifier was trained for 15 epochs, with both numbers of epochs chosen purposely as to limit over-fitting. Both were trained with a batch size of 10 segments, and both the encoder-decoder and the R-peak classifier models were trained using leave-one-subject-out cross validation.  

The path that the input signal takes through the combined model is highlighted in Fig.~\ref{model_walkthrough}. Observe that in this test example, where multiple peaks are present in the input with only true R-peak, the output of the ECG template ``matched filtering" layer results in 3 strong peaks. These peaks are then sifted through by the subsequent decoder layers to produce a single peak in the latent representation - a process we refer to as ``refinement". This peak in the latent representation is then used by the R-peak classifier to determine the true R-peak location. 

\begin{figure}[h!]
\centerline{\includegraphics[width=0.45\textwidth]{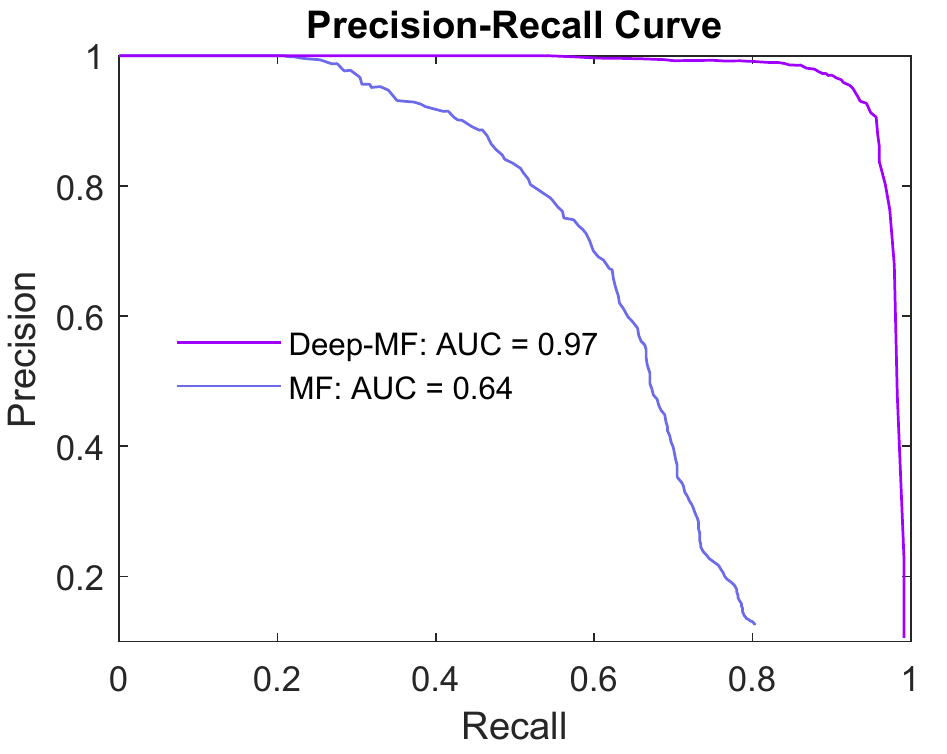}}
\caption{Precision-recall curves for R-peak detection, for the proposed Deep Matched Filter (Deep-MF) and the standard matched filter (MF). The Deep-MF (purple) achieves an area under the curve (AUC) value of 0.97, compared with the MF (blue) which achieves an AUC of 0.64.}
\label{recall_prec_curve}
\end{figure}

\subsection{Model Evaluation}

The deep matched filter (Deep-MF) was evaluated against two separate models, namely a standard matched filter (MF) and the matched filter Hilbert transform algorithm (MF-HT)\cite{Chanwimalueang2015}. Both MF and MF-HT were implemented using the input channel that was band-pass filtered between 1 and 45Hz. For the outputs of the Deep-MF and MF, R-peaks were determined using the MATLAB function findpeaks, with a maximum peak width of 25 samples and a minimum peak distance of 12 samples. The determined R-peaks were then compared to the true R-peaks, which were also calculated using findpeaks on the reference ECG signal. If the predicted peak was within 40ms of the true R-peak, it was considered a match. This condition was also applied to the output of the MF-HT. The proposed Deep-MF and the MF were both evaluated in terms of r-peak recall (the proportion of the R-peaks in the reference signal that were correctly identified) and r-peak precision (the proportion of predicted R-peaks which were true R-peaks). An area under the curve (AUC) value corresponding to a precision-recall curve was calculated for both the proposed Deep-MF and the standard MF, by varying the minimum peak height threshold of the findpeaks function. This precision-recall curve was generated with the median precision and recall values across all subjects. For the MF-HT algorithm, it was not possible to vary sensitivity in this way, and thus the implementation was compared Deep-MF and the standard MF with fixed threshold parameters that produced a good balance of recall and precision, 0.11 in the case of Deep-MF and 0.90 in the case of MF. Note that the large difference in threshold used between the Deep-MF and the standard MF stemmed from the fact that the outputs of the Deep-MF were scaled and thus lower in amplitude than the standard MF. The Deep-MF, standard MF and MF-HT were all compared again through performance in recall and precision, in the form of boxplots of these values across all 36 subjects. 

The effects of initialisation with an ECG template (shown in grey in Fig.~\ref{deep_learning_architecture}) were also evaluated and contrasted against random initialisation, both in terms of performance and interpretability. To evaluate the performance impact of initialisation, the mean absolute test error of the encoder-decoder model was calculated at regular intervals during training for both random initialisation and ECG template initialisation. Similarly, AUC values for R-peak recall-precision curves were calculated for both random initialisation and ECG template initialisation. For the purposes of interpretability, the kernel weights post training with ECG template initialisation were examined visually and compared to the initialised values, with a focus on the P, Q, R, S, and T portions of the ECG to determine which aspects of the ECG were valuable to the model for detecting ECG in the input.

\section{Results and Discussion}

The deep matched filter achieved a median R-peak detection recall of 94.9\% in the Ear-ECG of unseen subjects, with an interquartile range of 60.1\% to 99.3\%. The Deep-MF had a corresponding median precision of 91.2\% with an interquartile range of 68.6\% to 98.2\%. The high recall and precision of the Deep-MF is reinforced by an example Deep-MF model output shown in Fig.~\ref{results_mfht}(a), alongside the ground truth ECG and the input Ear-ECG. It can be observed in this example that even in a scenario with a poor signal-to-noise ratio, in which it is difficult to visually identify which peaks in the input belong to ECG, the Deep-MF correctly identifies the correct peaks and excludes the incorrect peaks. On the same subject pool, the MF-HT achieves a median recall of 83.4\% with an IQR of 62.3\% to 97.2\%. The MF-HT has a corresponding median precision of 79.5\% (IQR 28.6\% to 96.8\%). These results are comparable to the original implementation of MF-HT on noisy ECG by Chanwimalueang et al \cite{Chanwimalueang2015}, in which the algorithm achieved a recall of 83.1\% an a precision of 86.8\%. 

The results of the Deep-MF and MF-HT are compared with the standard MF which had a median recall of 62.3\% (IQR 27.3\% to 85.9\%) and a median precision of 67.1\% (IQR of 47.9\% to 84.5). The full results for the comparison of recall and precision between the Deep-MF, MF-HT and standard MF are shown in boxplots in Fig.~\ref{results_mfht}(b) with the Deep-MF plotted in purple, the MF-HT in red, and the standard MF in blue. Whilst the overall distribution of recall was comparable between the Deep-MF and MF-HT, with both models having a similar interquartile range, Deep-MF performed far better in terms of precision. This is likely due to the advantages of the Deep-MF having multiple refinement layers, in which false peaks could be discarded. It is important to note that the MF-HT relies on manual input to select a matched filter template from the input signal, which explains the improvements in recall over the standard-MF in which a fixed template was used across all subjects. Moreover, the MF-HT also has further conditions on determining which peaks are true R-peaks, based on a balance between the correlation between the template and the input and a deviation from the mean RR interval. These conditions explain why the median precision of the MF-HT was also higher than the standard MF.

For both the Deep-MF and standard MF, recall and precision were evaluated across the full range of peak sensitivity threshold values to produce precision-recall curves taken from the median results across all subjects, as shown in Fig.~\ref{recall_prec_curve}. The Deep-MF achieves an area under the curve value of 0.97, compared to the standard MF which achieves an AUC of 0.64. Observe that precision values never drop below 0.1 due to the two fixed conditions of this findpeaks implementation, namely the maximum peak width and a minimum peak distance, which limited precision from dropping below this value.

It is important to note that the Deep-MF model is higher in complexity than the MF-HT and the standard MF, which provides a more significant barrier to practical implementation. However, on the landscape of deep learning implementations, the Deep-MF is a relatively small model, with an implementation consisting of only 6 total layers, 5 of which are convolutional layers. Moreover, the total number of convolutional kernels is only 162, making the Deep-MF very computationally cheap to implement. As is the case with the standard MF and MF-HT, the Deep-MF can therefore operate in quasi real-time, with a rolling window of the previous 2 seconds of input data. Furthermore, the increase in model complexity of the Deep-MF is justified by vast improvements in performance, with a median increase in recall of 11.6\% and a median increase in precision of 11.8\% when compared to the MF-HT.

\begin{figure}[h!]
\centerline{\includegraphics[width=0.5\textwidth]{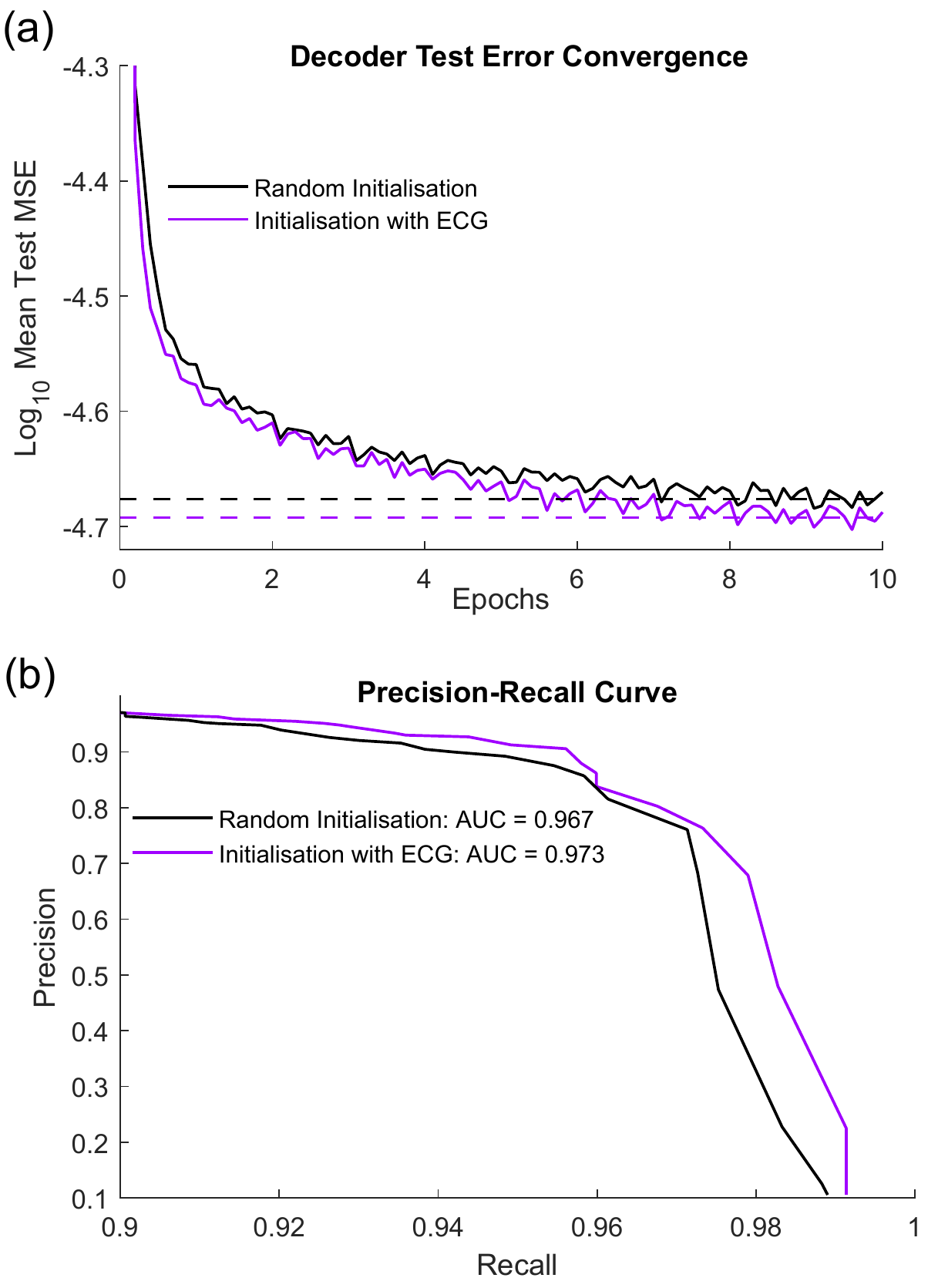}}
\caption{The effects on model performance of the random initialisation of kernel weights (black), against partial initialisation of the input ``Matched Filter" layer with an electrocardiogram template (purple). (a) The $Log_{10}$ of the test error convergence of the encoder-decoder module during the training process, with a dotted line representing the mean test error of the last half of the final training epoch. (b) Precision-Recall curves for R-peak detection of the Deep-MF, shown to have an area under the curve (AUC) value of 0.967 with random initialisation and an AUC value of 0.973 after initialisation with an ECG template.}
\label{initialisation_accuracy}
\end{figure}

\begin{figure}[h!]
\centerline{\includegraphics[width=0.45\textwidth]{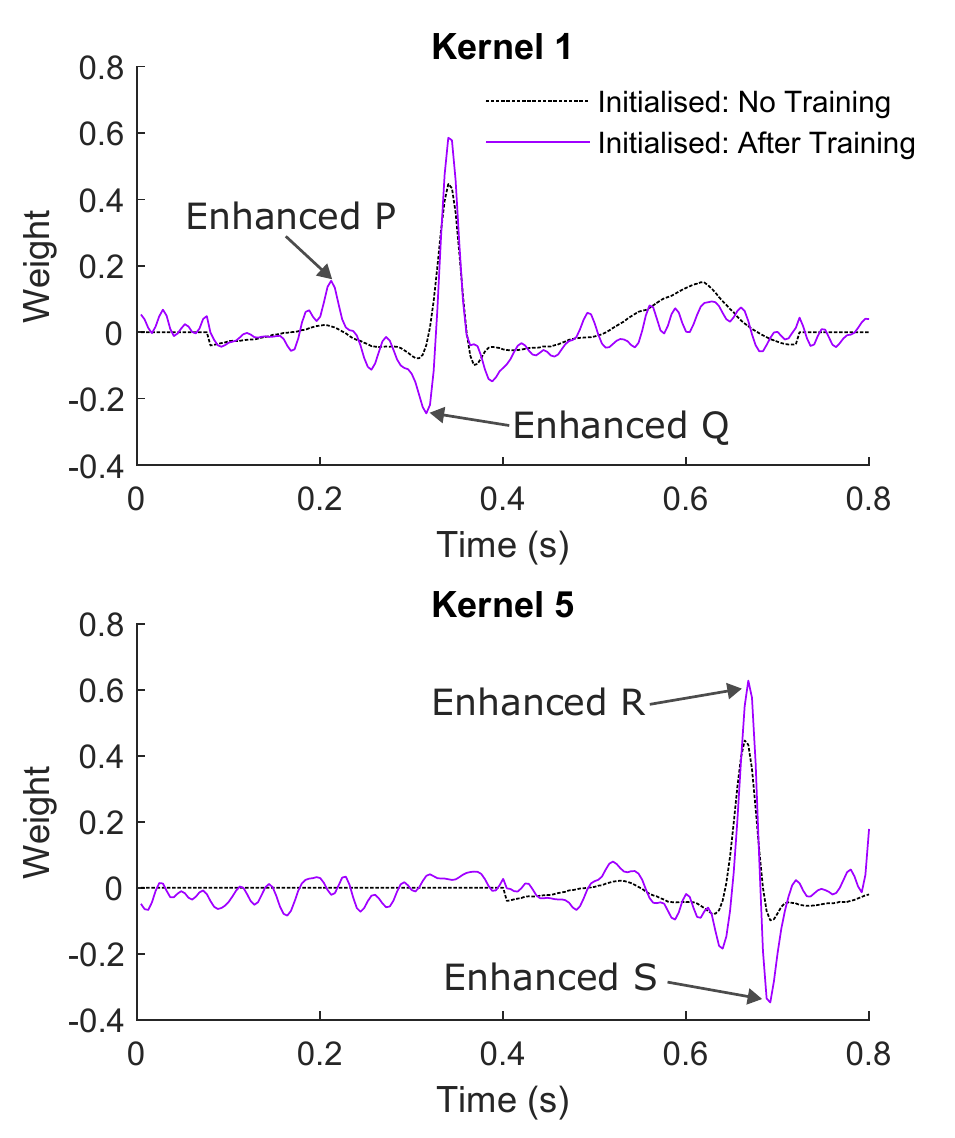}}
\caption{The effects of training on kernels initialised with an electrocardiogram template. The dashed trace in black represents the initialised ECG kernel weights before the training process, and the purple trace represents the kernel weights after the training process. Labelling highlights that in different kernels, different aspects of the ECG are amplified during the training process. In the 1\textsuperscript{st} kernel (top), there is an enhancement of the P wave and of the Q portion of the QRS complex. In the 5\textsuperscript{th} kernel there is an amplification of the S portion of the QRS complex. In all kernels there is an amplification of the R-peak that was provided by the ECG template.}
\label{kernel_plot}
\end{figure}

\section{Interpretability}

A major barrier to the widespread adoption of deep learning techniques in digital health is the notion that the models themselves are a black box that offer no interpretability to explain the predictions they make. In this paper, we have shown that this particular model for R-peak detection is fully interpretable as a multi-layer matched filter, a tool for detecting an overlap in patterns between an input signal and a template of interest. A key aspect of this argument is the partial initialisation of the input layer of the Deep-MF to the pattern which it is trying to detect, namely the electrocardiogram. For rigour, it is important to also examine the same kernels after the training process, as if the network was to completely discard the ECG template then it could be assumed that the model did not find the information useful for minimisation of error and thus was not searching for ECG ``matches" in the input signal.

In terms of model accuracy, the effects of partial initialisation of the input layer of the Deep-MF with the same ECG template as used in the standard matched filter were small. It is highlighted in Fig.~\ref{initialisation_accuracy}(a) that initialisation with an ECG template provided a minor improvement in the mean squared error of the decoder output, and in Fig.~\ref{initialisation_accuracy}(b) it is shown that this corresponded to a slight increase in precision-recall AUC from 0.967 to 0.973. Note that the precision-recall curves plotted in Fig.~\ref{initialisation_accuracy}(b) are zoomed in the recall axis to exaggerate the difference between the random initialisation and the template initialisation. Whilst these improvements in performance are marginal, it does suggest that initialising the Deep-MF with an ECG provided the model with useful information that it didn't otherwise learn from the training process.

When examining the effect of training on the initialised weights, as shown with two examples in Fig.~\ref{kernel_plot}, it is clear that the network holds on to aspects of the ECG templates as it deems them useful in minimising error. Notably, in the all kernels initialised with an ECG template, the R-peak information from the template was retained by the network and exploited with an increase in the weights at this location. Moreover, the network goes further than the original template ECG and exaggerates aspects such as the P wave, and the Q and S parts of the QRS complex during the training process, showing that these aspects of the ECG are useful in distinguishing true R-peaks from other peaks in the input signal. This can be seen in Fig.~\ref{kernel_plot} with the kernel 1 seeing an amplification in weights around the P wave and Q, and kernel 5 seeing an amplification of the R and S components of the ECG.

\section{Conclusion}

We have introduced a novel Deep Matched Filter framework for the detection of R-peaks in wearable-ECG. The proposed Deep Matched Filter (Deep-MF) has been evaluated on the Ear-ECG of 36 subjects, and has shown a marked improvement over existing matched filter based algorithms, both in terms of recall and precision. In parallel with demonstrating the proficiency of the Deep-MF at R-peak detection in scenarios with poor signal to noise ratio, it has been illustrated that this encoder-based model behaves precisely as a learned matched filter. It serves to detect ECG segments in the input, followed by several refinement layers which distinguish the true ECG matches from the false matches. This has been reinforced through partial initialisation of the model with an ECG template, whereby through the training process the model amplifies physically relevant aspects of the ECG. The proposed Deep Matched Filter has been shown to greatly improve the practical utility of the Ear-ECG signal, whilst being transparent in its operation. It is our hope that physically grounded models such as the Deep Matched filter may help to accelerate wide scale adoption of interpretable artificial intelligence in healthcare.

\section*{Acknowledgment}
This work was supported by the USSOCOM MARVELS grant and the Dementia Research Institute at Imperial College London.

\FloatBarrier
\bibliographystyle{IEEEtran}
\bibliography{IEEEabrv,deep_matched_filter_ecg}

\end{document}